# CONTROL COMPENSATION BASED ON UPPER BOUND DELAY IN NETWORKED CONTROL SYSTEMS

Vatanski[b] N., Georges J.-P.[a], Aubrun C[a]., Rondeau[a] E. and S.-L. Jämsä-Jounela[b]

[a]) *Research Centre for Automatic Control (CRAN UMR 7039) Henri Poincaré University, France*
[b]) *Laboratory of Process Control and Automation, Helsinki University of Technology, Finland*
E-mail: Nikolai.Vatanski@hut.fi, Jean-Philippe.Georges@cran.uhp-nancy.fr, christophe.aubrun@cran.uhp-nancy.fr, Eric.Rondeau@cran.uhp-nancy.fr, sirkka-l@hut.fi

Abstract: Recent interest in networked control systems (NCS) has instigated research in both communication networks and control. Analysis of NCSs has usually been performed from either the network or the control point of view, but not many papers exist where the analysis of both is done in the same context. In this paper an overall analysis of the networked control system is presented. First, the procedure of obtaining the upper bound delay value for packet transmission in the switched Ethernet network is presented. Next, the obtained delay estimate is utilised in delay compensation for improving the Quality of Performance (QoP) of the control systems. The presented upper bound delay algorithm applies ideas from network calculus theory. For the improvement of QoP, two delay compensation strategies, the Smith predictor based and the robust control based delay compensation strategies, are presented and compared.

Keywords: Networked control systems, delay compensation, real time systems, network calculus,

## 1. INTRODUCTION

Automation systems of the future, and even those currently in use today, will consist of a large number of intelligent devices and control systems connected by local or global communication networks. In these networked control systems (NCSs), communication between process, controllers, sensors and actuators is performed through the network. The primary benefits in developing the systems from point-to-point systems towards NCS like systems, are reduced system wiring, ease of system diagnosis and maintenance, and increased system agility.

In most cases insertion of the network does not significantly affect the performance of the control system. However, for some real-time processes, care should be taken when implementing a NCS. For such processes, the insertion of the communication network into the feedback control loop introduces an additional, either constant or time varying delay, that makes the analysis and design of the NCS more complex.

There are the three main directions in approaching the problem of network-induced delays in NCS. One way is to design a controller irrespective of the delay, and then to design a network scheduling procedure so that the delay is minimized. The second approach is to study the NCS problem as an integration of network and control design. This paper addresses the third approach in which the control strategy is designed so that it compensates a priori the networked-induced delay. During the past few years this topic has been actively researched and several compensation strategies have been proposed. Extensive state of the art articles and surveys have been published (see Tipsuwan and Chow, 2003; and Richard, 2003). The delay compensation methodologies proposed apply ideas from the following control theory fields: robust control (Göktas, 2000), LQG-optimal control (Nilsson, 1998), LMI based control (Li *et at.*, 2004). More specific strategies include: fuzzy logic based control (Almutairi *et al.*, 2001), gain adaptation of controllers (Tipsuwan and Chow, (2002), Smith predictor based compensation (Bauer *et al.*, 2001), to name but a few.

However, in these papers the assumption has usually been made that information about the network effect (delay distribution, uncertainty, deviation from mean value, missing value rate) is known in advance, and that the information is used in the design or synthesis of the control law. The whole procedure of obtaining information about the network delay and using it in control system design and synthesis is given in a few papers. This is the estimation of the network properties, and using these in control compensation is still usually performed in networking and control communities separately.

This paper addresses the gap that still exists between the two communities. In this paper the procedure of obtaining information about the delay (the upper bound) is presented and the obtained value is utilised in delay compensation.

The delay algorithm presented applies ideas from network calculus theory, and the delay compensating strategies are based on the Smith predictor and robust control theory.

The paper is organised as follows: Chapter 2 is dedicated to introducing the upper bound delay estimation algorithm. In Chapter 3 the delay compensation strategies are introduced. Chapter 4 consists of the simulation results and discussion, And the paper ends with a concluding section in Chapter 5.

## 2. UPPER BOUND DELAY ESTIMATION

In this paper the switched Ethernet network is used as an example of the NCS network. The Ethernet networks are nowadays also more and more used in control applications and, in this context, it is important to understand the behaviour of the network in order to be able to control the network performance, such as delays (Georges *et al.*, 2004).

Next, the procedure of obtaining the upper bound delay over the network will be explained in more detail. How to obtain a maximum delay for crossing a single Ethernet switch will be explained in Section 2.1, and the procedure of obtaining end-to-end delays in the network, based on the delays over the switches, will be given in Section 2.2.

The communication network upper bound delay estimation algorithm presented in this paper applies ideas from network calculus theory (see Cruz, 1991; Le Boudec and Thiran, 2001; Jasperneite *et al.*, 2002). For more details of the algorithm, see Georges *et al.* (2005). The communication network is represented as a network of interconnected switches, and each switch is modelled as a combination of the basic components: multiplexers, demultiplexers and FIFO queues, see Fig. 1.

### 2.1 Maximum delay for crossing the Ethernet switch

To obtain the upper bound delay for crossing a single Ethernet switch, the upper bound delays over the basic components should first be determined. In this section we will show how to obtain the upper bound delay for the FIFO multiplexer, FIFO queue, and demultiplexer basic components. The upper bound delay over the switch is then the sum of the upper bound delays over the basic components:

$$\overline{D_{switch}} = \overline{D_{mux}} + \overline{D_{queue}} + \overline{D_{output}} \quad (1)$$

The traffic arriving at the switch, both periodic and aperiodic, is modelled as a "leaky bucket controller". That is, the data will arrive at the switch only if the level of the amount of data in the buffer of the switch is less than the maximum buffer size and the data leaves the switch at a constant rate.

*Upper bound delay over a FIFO multiplexer.* The first step in calculating the delay over a multiplexer is to determine the arrival curves of the traffic coming to the component and the service curves provided by the component. With the assumption that the traffic follows the leaky bucket controller, these curves are affine and have the form of:

$$b(t) = \sigma + \rho t \quad (2)$$

Where $\sigma$ is the maximum amount of data that can arrive in a burst, and $\rho$ is an upper bound of the average rate of the traffic flow. Typical arrival and service curves are shown in Fig. 2.

The next step is to determine the upper bound backlog in the multiplexer. The backlog is the number of bits accumulated in the component, and it is a measure of congestion over the component. For the arrival and service curves in Fig. 2, the upper bound backlog occurs at time $t$ and can be calculated from

$$b_1(t) + b_2(t + L/C_2) - C_{out}t \quad (3)$$

Where $b_1$ and $b_2$ are the arrival curves of stream 1 and 2 at time $t$, $L$ is the maximum length of the frames, $C_2$ is the capacity of the import port 2, and $C_{out}$ is the capacity of the output link.

When the upper bound backlog over the component is known, the upper bound delay over the multiplexer component is then obtained by dividing the maximum backlog value by the capacity of the output link of the multiplexer.

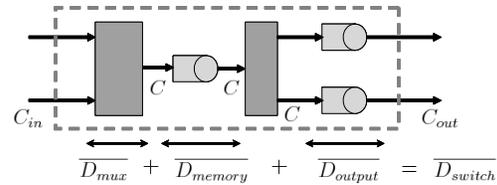

Fig. 1. Model of a 2 port-switch in a full duplex mode based on shared memory and a cut-through management.

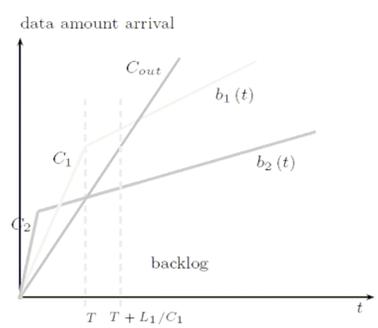

Fig. 2. Arrival and service curves and backlog evolution inside the two-input FIFO multiplexer.

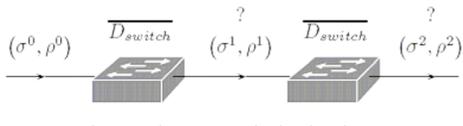

Fig. 3. Burstiness along a switched Ethernet network.

The procedure can be summarized as follows:

In a FIFO m-inputs multiplexer, the delay for any incoming bit from the stream *i* is upper-bounded by:

$$\overline{D_{mux,i}} = \frac{1}{C_{out}} \min_k \overline{B_{mux,k}} \quad (4)$$

Where $\overline{B_{mux,k}}$ is an upper-bound of the backlog in the bursty periods $u_k$.

For $k = i$ (i.e. $b_i$ is bigger than $b_k$), the bursty period is defined by $u_i = \sigma_i /(C_i - \rho_i)$ and the backlog is upper-bounded by:

$$\overline{B_{mux,i}} = \sum_{z=1; z \neq i}^{m} \left( \sigma_z + \rho_z \left( u_i + \frac{L_z}{C_z} \right) \right) + u_i (C_i - C_{out}) \quad (5)$$

Where $\sigma_i$ is the burstiness of the stream *i*, $\rho_i$ is the average rate of arrival of the data of stream *i*, $L_i$ is the maximum length of the frames of stream *i* and $C_i$ is the capacity of the import port *i*.

For $k \neq i$ (i.e. $b_i$ is smaller than $b_k$) such that $1 \leq k \leq m$, we have $u_k = \sigma_k /(C_k - \rho_k) - L_k / C_k$ and

$$\overline{B_{mux,i}} = \sum_{z=1; z \neq k}^{m} \left( \sigma_z + \rho_z \left( u_k + \frac{L_z}{C_z} \right) \right) + u_k (C_k - C_{out}) - \rho_i \frac{L_i}{C_i} + L_k \quad (6)$$

*Upper bound delay over a FIFO queue.* For the FIFO queue the delay of any byte is upper-bounded by:

$$\overline{D_{queue}} = \frac{1}{C_{out}} \frac{(C_{in} - C_{out})}{C_{in} - \rho_{in}} \sigma_{in} \quad (7)$$

*Upper bound delay over a demultiplexer.* The demultiplexer has one input link and two or more output links. Its function is to split the streams that arrive at the input ports and to route them to the appropriate output ports. In Ethernet, this consists of reading the destination address at the start of the frame and selecting the output port associated with its destination in the forwarding table. Due to the Spanning Tree Protocol, only one path is activated to go from one point to another. Therefore it is assumed that the routing step is instantaneously achieved. Thus the demultiplexer does not generate delays.

*2.2. Maximum end-to-end delays for crossing a switched Ethernet network*

The upper-bounded delay equations for crossing a switch were proposed in the previous section. In the equations, the maximum delay value $\overline{D}$ depends on the leaky bucket parameters: the maximum amount of traffic $\sigma$ that can arrive in a burst and the upper bound of the average rate of the traffic flow $\rho$. In order to calculate the maximum delay over the network, it is necessary that the envelope $(\sigma, \rho)$ is known at every point in the network. However, as shown in Fig. 3, only the initial arrival curve values $(\sigma^0, \rho^0)$ are usually known, and the values for other arrival curves should be determined. To calculate all the arrival curve values the following equations can be used:

$$\sigma_{out} = \sigma_{in} + \rho_{in} D$$
$$\rho_{out} = \rho_{in} \quad (8)$$

For example, for the arrival curve $(\sigma^1, \rho^1)$ in Fig. 3 the envelope after the first switch is:

$$(\sigma^1, \rho^1) = (\sigma^0 + \rho^0 \overline{D_{switch}}, \rho^0) \quad (9)$$

Now it possible to summarize the procedure of obtaining the maximum end-to-end delays in a switched Ethernet network. The algorithm is the following:

1. Identify all streams on each station and determine the initial leaky bucket values.
2. Identify the route of each stream. In the switched Ethernet networks, the paths are determined by the spanning tree protocol.
3. On each switch, formulate the output burstiness equations for all streams.
4. Define the equation systems of form $A\Psi = \Phi$ or $a_n \sigma_1 + b_n \sigma_2 + ... + z_n \sigma_m = \delta_n$
5. Solve the burstiness values.
6. Determine the end-to-end delay in the network from the equation

$$\overline{D_i} = \frac{\sigma_i^h - \sigma_i^o}{\rho_i} \quad (10)$$

where *h* is the number of crossed switch.

## 3. DELAY COMPENSATION USING THE UPPER BOUND DELAY ESTIMATE

In the NCS environment the main goal of the control system is to maintain Quality of Performance (QoP) of the control system regardless of the delays in the network. The system should be robust and be able to compensate the delay induced by the network. Prior to presenting the delay compensation strategies it is important to state the following assumptions about the process and the network:

1. In the network all control and measurement information is sent in a single packet.
2. The process is assumed to be fast. The sampling time necessary to capture all relevant process dynamics is significantly smaller than the network induced delay. (If the sampling time necessary to capture all process dynamics is larger than the network induced delay, then the delay will have a similar effect on QoP of the control system as a small additional measurement error.

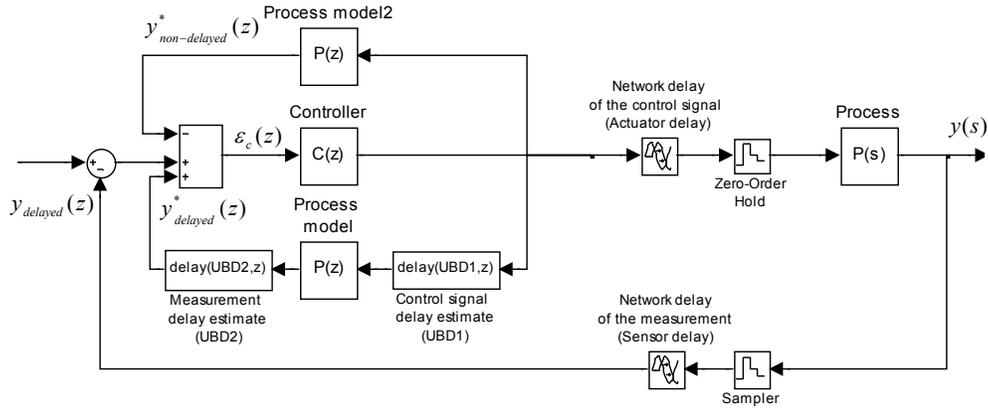

Fig. 4. A Smith predictor scheme for compensation of the network induced delay

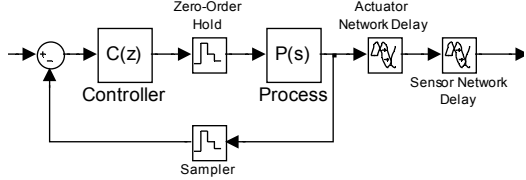

Fig. 5. The equivalent Smith predictor compensation scheme in case the delay is commutative

3. No packet losses occur in the communication network.

In the following sections the two delay compensation strategies will be presented. First, the compensation strategy based on the Smith predictor will be introduced, and next the robust control theory based approach will be presented. In order to facilitate understanding, the synthesis of both is done in continuous time. For implementation the obtained controllers should be discretized. If assumption (2) above holds, the results for the discrete case will be similar.

*3.1 Smith predictor based approach*

The delay compensation scheme of the Smith predictor is shown in Fig. 4. In the figure, minor feedback loops have been introduced around the conventional controller.

In order to see how the scheme works, let us proceed to analyze the time-delay compensator assuming that there are no model errors in the scheme (process model and a time delay are known exactly) and that the time delays are commutative. This means that the delayed process measurement and the delayed process output estimate are equivalent, $y_{delayed} = y^*_{delayed}$. Then observe that the signal reaching the controller is a corrected error signal given by:

$$\varepsilon_c = r - y_{delayed} + y^*_{delayed} - y^*_{non-delayed} \quad (11)$$

$$\text{or } \varepsilon_c = r - y^*_{non-delayed} \quad (12)$$

That is the error signal that reaches the controller is calculated on the basis of the non-delayed estimate of the process output. Implying, as a result, that the block diagram in Fig. 4 is equivalent to that shown in Fig. 5. The net result of introducing minor loops is therefore to eliminate the time delay factor from the feedback loop where it causes stability problems, and move it outside the loop where it has no effect on closed-loop stability.

The scheme works well as long as the process model and the time delay are known. In the case of modeling errors, the performance of the Smith predictor decreases. In addition, the Smith predictor scheme is designed for constant time delays and therefore may not perform as well for systems with time delays that significantly vary over time.

To increase the robustness of the Smith predictor so that it is more suitable for the NCS environment where time delay varies, we use the following delay estimate obtained on the basis of the upper-bound delay value. This delay estimate was originally proposed by Wang, *et al.*, (1994) to represent uncertain delays in the $H_\infty$ framework in the design of robust controllers.

In continuous time, the delay can be approximated by:

$$e^{-\tau\Delta s} \approx 1 - \frac{UBD \cdot s}{1 + \frac{UBD}{3.645}s}\Delta, |\Delta| \leq 1 \quad (13)$$

Where $\Delta$ describes uncertainty, and *UBD* is the upper bound delay.

*3.2 Robust control theory based approach*

The robust control approach is used as a second delay compensation strategy. The main idea is the following: first, the network-induced delays are represented in the frequency domain as an uncertainty around the nominal plant. Next, using the robust control methods ($H_\infty$-synthesis, D/K iteration etc.) a controller is generated that enables maintenance of the QoP of the control system, even in a worst case disturbance. In this case the worst case performance is when the network delay corresponds to the upper bound delay.

Compared to the Smith predictor based approach, the benefits of this approach are the following. The robust control approach is based on the worst case uncertainty, thus no information is needed about the distribution of the delay. In addition, the uncertainties about the process, as well as about the network, can be handled using the same methodology. For the process the uncertainty about a gain, time constant, pole and zero location, and for the network uncertainty induced by the network delay, jitter and the effect of missing values, can be handled using the same methodology.

*Representing the network induced delay.* The network delay can be represented as a multiplicative uncertainty around the plant:

$$G_p(s) = G(s)(1 + w_I(s)\Delta(s)); \underbrace{|\Delta_I(jw)| \leq 1 \forall}_{\|\Delta_I\|_\infty \leq 1} \quad (14)$$

Where $w_I$ is a weight used to describe the delay uncertainty. The weight can be obtained by finding the smallest radius $l_I(w)$ that includes all possible plants:

$$l_I(w) = \max_{G_p \in \Pi} \left| \frac{e^{UBD_1 s} G(jw) e^{UBD_2 s} - G(jw)}{G(jw)} \right| = \max_{G_p \in \Pi} \left| e^{UBD_1 s} e^{UBD_2 s} - 1 \right|$$

$$l_I(w) = \begin{cases} \left| e^{(UBD_1 + UBD_2)s} - 1 \right|; w < \pi/(UBD_1 + UBD_2) \\ 2; w > \pi/(UBD_1 + UBD_2) \end{cases} \quad (15)$$

And choosing the weight $w_I$ such that

$$|w_I(jw)| \geq l_I(jw); \forall w \quad (16)$$

For example in this case the following weight can be chosen:

$$w_I(s) = \frac{(UBD_1 + UBD_2) \cdot s}{1 + (UBD_1 + UBD_2) \cdot s / 3.465} \quad (17)$$

*Controller synthesis.* For a SISO case the controller synthesis problem can be solved in a relatively straightforward manner, since the SISO case with one complex multiplicative perturbation the Robust Performance (RP) problem can be approximated as a weighted mixed sensitivity problem where the condition is slightly strengthened:

$$\left\| \begin{matrix} w_P S \\ w_I T \end{matrix} \right\|_\infty = \max_\omega \sqrt{|w_P S|^2 + |w_I T|^2} < \frac{1}{\sqrt{2}} \quad (18)$$

Where $w_P$ is a weight for the sensitivity function $S$ (usually an approximator of an integrator), and $T$ is the complementary sensitivity function.

For a MIMO case (or for a SISO with additive uncertainties) the use of a more complicated technique such as $\mu$ synthesis is required.

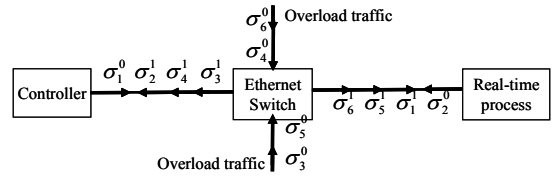

Fig. 6. The structure of the network

## 4. SIMULATION RESULTS AND DISCUSSIONS

In the simulations, the network of a real time process, a controller, and two overload traffic stations connected over a full duplex Ethernet switch, were used. The structure of the system is shown in Fig. 6.

To calculate the upper bound delay, the initial leaky bucket values of each stream were first identified. 6 are messages sent periodically. The traffic sent from the process to the controller is given by $b_1^0(t)$, and the traffic from the controller to the process by $b_2^0(t)$. The upper-bounds for these traffics will be computed in order to obtain the upper bounds, $UBD_1$ and $UBD_2$. We consider also background traffic ($b_3^0(t)$, $b_4^0(t)$, $b_5^0(t)$, $b_6^0(t)$) from the stations to the process and to the controller in order to overload the network:

$$b_1^0(t) = b_2^0(t) = \sigma_1^0 + \rho_1 t = 72 + 7200t$$
$$b_3^0 = b_4^0(t) = b_5^0(t) = b_6^0(t) = \sigma_3^0 + \rho_3 t = 1526 + 305200t \quad (19)$$

Next, the route of each stream was identified and the output burstiness equations were formulated. After solving the burstiness values the end-to-end upper bound delay for streams 1 and 2 are:

$$UBD_1 = UBD_2 = \frac{\sigma_1^2 - \sigma_1^0}{\rho_1} \approx 3.5\,ms \quad (20)$$

In evaluating the effects of the network on the control system performance, the following model of a real time process and a nominal controller were used (time in *ms*):

$$P(s) = \frac{2}{(s+5)(s+0.2)} \quad (21)$$

$$C(s) = \frac{K_P s + K_I}{s}, \quad K_P = 0.5508, \quad K_I = 0.4529$$

The controller parameters were obtained by minimizing the integral of the square errors (ISE) for the system with the network delay in an actuator and sensor paths of 1 ms.

First, the delay compensation strategy based on the Smith predictor presented in Fig. 4 was used. The model was assumed to be known, and the network delay in a sensor and in actuator sides were assumed to vary randomly between zero and the upper delay value estimate. Equation 13 was used for both the measurement delay estimate and the control signal delay estimate.

Next, the delay compensation strategy based on the robust control approach was implemented by solving the mixed sensitivity problem in Equation 18. Equation 17 was used as a weighting function for the complementary sensitivity function $T$. As a weighting function $w_P$ for the sensitivity function $S$ the following approximation of the integrator was implemented:

$$w_P(s) = \frac{s/M + \omega_B}{s + \omega_B A} \qquad (22)$$

Where $\omega_B$ is the bandwidth where control is effective, M is the desired maximum peak of wb, and A is a small number used to avoid numerical problems. The $H_\infty$ optimal controller for this mixed sensitivity problem was found using the Matlab$^{TM}$ Robust control toolbox.

The simulation results are presented in Fig. 8. Four graphs are shown: the system performance under a nominal controller (Equation 21) when the network delay is negligible, the system performance under a nominal controller when there is no delay compensation, the system performance when the Smith predictor is introduced, and the system performance under a robust controller. From the figure it can be concluded that the nominal system becomes unstable when the delay increases. The stability of the feedback control loop can, however, be regained even when a delay compensation strategy such as the Smith predictor is implemented. The performance of the control system can be improved by using the more advanced delay compensation/toleration procedure.

## 6. CONCLUSION

In this paper an analysis of the networked control system has been presented. A procedure for obtaining the upper bound delay value in the switched Ethernet network was presented, and the obtained delay estimate was used in the control compensation. Two control compensation strategies, the Smith predictor based compensation strategy and the robust control based compensation strategy, were presented and compared. It can be concluded that the upper bound delay estimate is an important measure of the networked control system which can also be used for the design and synthesis of a control system.

## ACKNOWLEDGEMENTS

This research has been conducted as a part of the Networked Control Systems Tolerant to Faults (NeCST) project IST-004303 that is partially funded by the EU. The authors gratefully acknowledge the support.

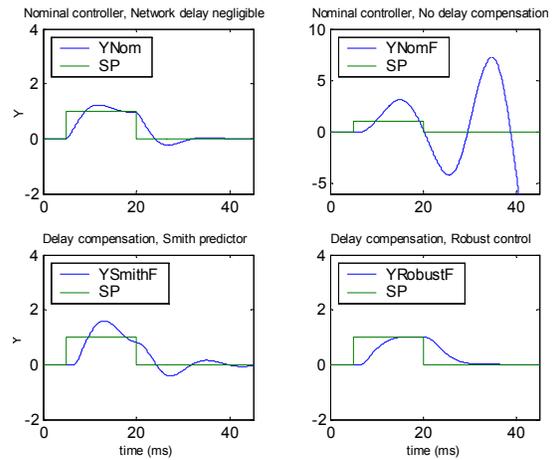

Fig. 8. The performance of the control system under various delay compensation strategies